# Integrated Design of Aluminum-Containing High-entropy Refractory B2 Alloys with Synergy of High Strength and Ductility

Jie Qi[1], Xuesong Fan[2], Diego Ibarra Hoyos[1], Michael Widom[3,4], Peter K. Liaw[2], and Joseph Poon[1,5]

[1] Department of Physics, University of Virginia, Charlottesville, VA 22904

[2] Department of Materials Science and Engineering, The University of Tennessee, Knoxville, TN 37996

[3] Department of Physics, Carnegie Mellon University, Pittsburgh, PA 15213

[4] Department of Materials Science and Engineering, Carnegie Mellon University, Pittsburgh, PA 15213

[5] Department of Materials Science and Engineering, University of Virginia, Charlottesville, VA 22904

**Abstract**

Refractory high-entropy-alloys (RHEAs) are promising high-temperature structural materials. Their large compositional space poses great design challenges for phase control and high strength-ductility synergy. The present research pioneers using integrated high-throughput machine learning with Monte Carlo simulations to effectively navigate phase-selection and mechanical-properties predictions, developing aluminum-containing RHEAs in single-phase ordered B2 alloys demonstrating both high strength and ductility. These aluminum-containing RHEAs achieve remarkable mechanical properties, including compressive yield strengths up to 1.6 GPa, fracture strains exceeding 50%, and significant high-temperature strength retention. They also demonstrate a tensile yield strength of 1.1 GPa with a tension ductility of 6.3%. Besides, we identify a valence-electron-count domain for alloy brittleness with the explanation from density-functional theory and provide crucial insights into elements' influence on atomic ordering and mechanical performance. The work sets forth a strategic blueprint for high-throughput alloy design and reveals fundamental principles that govern the mechanical properties of advanced structural alloys.

## 1. Introduction

In the ceaseless quest to defy extreme temperatures and hostile environments, high-performance alloys become indispensable in aerospace, automotive, and power generation sectors. Conventional Ni/Co-based superalloys, despite their thermal stability and high-temperature mechanical properties, have inevitably encountered inherent performance ceilings, such as a melting point below 1,500°C. The past decade has seen the advent of High-Entropy Alloys (HEAs) [1] with revolutionized performance. The refractory high-entropy alloys (RHEAs), comprising refractory elements, Ti, V, Cr, Zr, Nb, Mo, Hf, Ta, and W, have stood out for high-temperature applications with high strength and ductility [2,3]. However, their broader adoption faces challenges, such as high cost, densities, and poor oxidation resistance. Al-containing refractory high-entropy alloys (Al-RHEAs) resembling aluminides have emerged as a solution [4–7], with the Al incorporation yielding cost/density reductions, and enhanced oxidation and corrosion resistance [8,9]. Moreover, Al inclusion typically induces the formation of an ordered body-centered-cubic (BCC)-derivative (B2) phase within the disordered BCC matrix [10–12], further strengthening the alloys.

Despite the noteworthy advantages of Al-RHEAs, their design inevitably introduces specific challenges. Most Al-RHEAs demonstrate limited ductility. The strong p-d electron interaction between Al and refractory elements can precipitate brittle intermetallic phases (IMs) [11,13]. The dislocation movement impediment from B2 long-range ordering (LRO), and the reduced slip planes from the BCC to B2 phase, further limit their ductility [10]. Therefore, careful phase and B2-LRO control becomes imperative. Besides, the design challenge resides in effectively navigating the vast HEA compositional space and optimizing manifold materials' properties simultaneously to achieve the Pareto Front. The conventional trial-and-error approaches have become obsolete, making Integrated-Computational-Material-Engineering (ICME) vital in material exploration [14–17]. Moreover, how various material parameters fundamentally influence alloy formation and ductility are yet to be fully elucidated, which remains an area of significant research focus.

The novelties of the present work are three core innovations that address the aforementioned challenges: First, a novel alloy-design strategy merges the cutting-edge Machine Learning (ML) models enhanced ICME framework with Monte Carlo (MC) simulations and classical computational methodologies, enabling high-throughput HEA predictions, pinpointing potential high-performance alloys, and reducing experimental efforts. Second, joint endeavors from experiments and MC studies successfully optimize a series of B2-phase AlHfNbTi(V) HEAs, exhibiting superior strength and ductility, compared to typical plasticity-constrained B2 Al-RHEAs, highlighting aluminum's significance in atomic ordering and plasticity control. Third, a prominent correlation between the valence electron count (VEC) [18] and alloy ductility elucidated in light of Density Functional Theory (DFT) calculations underscores VEC's importance within the property-prediction paradigm.

Moving forward, this article will provide insights into a high-performance HEAs development methodology via ICME and theory-guided efficient compositional tuning. It delves into fundamental phenomena regarding elements' influence on atomic ordering and the role of electronic structures on HEA-mechanical properties, laying a foundation for future alloy development and inspiring research in this field.

## 2. Results

**ICME Model Construction and Results**

The ICME model (Figure 1) integrates phases and properties prediction modules to design

Al-RHEAs with high strengths and ductility. Considering phase predictions, CALPHAD (CALculation of PHAse Diagrams) frequently identifies the Al-RHEAs' Al-X-Y type (X and Y are refractory elements) B2-strengthening phase [19] as the BCC matrix [19,20]. Therefore, we adopt ML models trained specifically for Al-X-Y B2 and other phases (detailed in our prior work [21,22]), which integrate innovatively designed phase-diagram features [21] with traditional Hume-Rothery and thermodynamic features [23], and are enhanced by feature engineering for improved accuracy. Trained on a database of ~1,000 HEAs and comprehensively validated by experiments, the ML models can predict nine phases with around 90% accuracy. Our targeted phases allow a solid-solution matrix and especially B2 phase in an as-cast condition, avoiding undesirable Laves, Sigma, or Heusler phase.

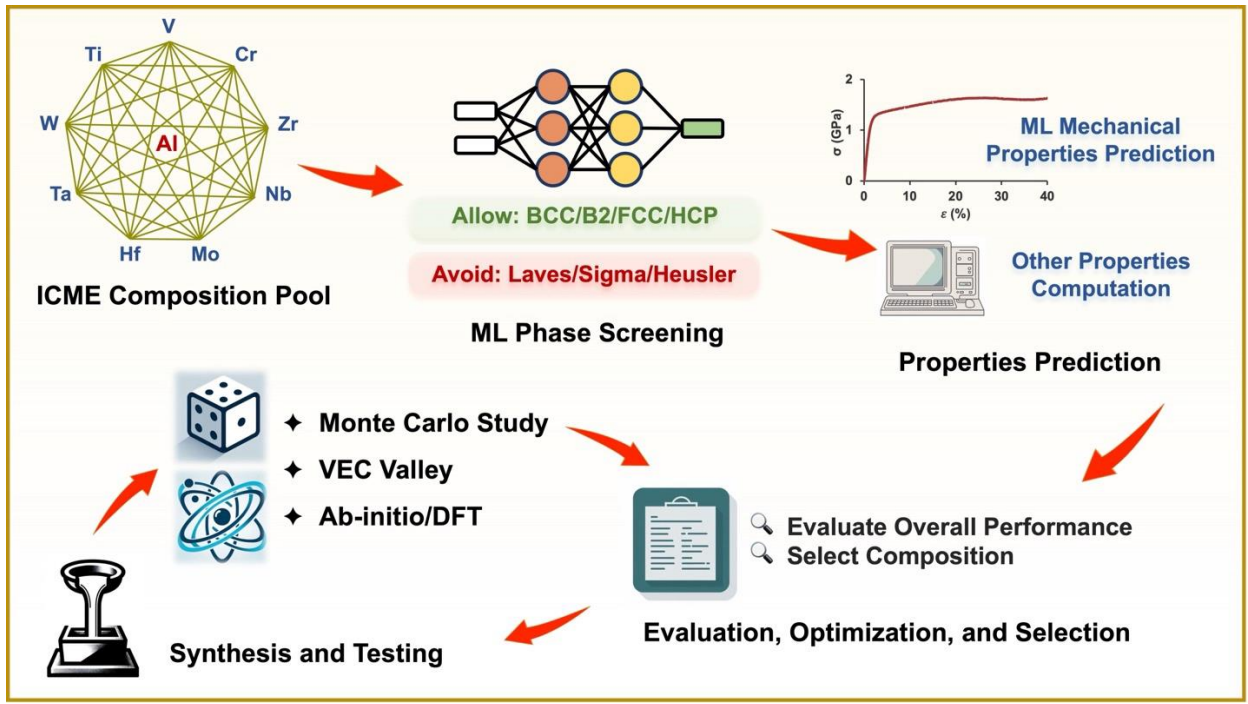

Figure 1. Schematic diagram showing the Al-RHEAs-design process. The abbreviations, ICME, ML, VEC, and DFT, denote Integrated Computational Material Engineering, Machine Learning, Valence Electron Count, and Density Functional Theory, respectively.

Properties predictions utilize ML-based mechanical-properties model to obtain compressive yield strength, $\sigma_{YS}$, and fracture strain, $\varepsilon_f$. Other properties including melting temperature, $T_{melt}$, density, $\rho$, and Poisson's Ratio, $\bar{v}$ are computed separately. The mechanical properties model utilizes thirteen initial physics-based features (Table S1), with some computed by the Effective Medium Calculation (EMC) method [24] to reduce the computational burden while maintaining fidelity to experimental or DFT-calculated values. The EMC's effectiveness is exemplified in the D parameter calculation [25] (methods section), whose DFT-calculated values show a direct correlation with the BCC HEAs' $\varepsilon_f$ [25,26]. The EMC-computed D parameter is highly aligned with the DFT-calculated values [25] with a mere 5% absolute mean error (Figure 2a), demonstrating a fast and reliable calculation alternate method for high-throughput and ICME applications.

Most current ML-based mechanical properties models are tailored to specific HEA phases to enhance prediction accuracies by focusing on phase-specific properties controlling mechanisms. The ICME-designed Al-RHEAs predominantly form an Al-X-Y B2 phase [19]. However, due to the limited mechanical-properties data available for B2 Al-RHEAs, we alternatively develop a baseline ML model using more abundant disordered BCC RHEAs and Al-RHEAs. Because the B2 phase's LRO can introduce ordering-strengthening, enhancing strength and reducing ductility [25,27], the baseline ML-model based on BCC HEAs might underestimate strength and overestimate the ductility of B2 HEAs. Our goal is to identify Al-RHEAs with inherent ductility predicted in their BCC phase, as brittle BCC Al-RHEAs are unlikely to manifest ductility upon the formation of B2-LRO. Furthermore, 30 % of the ML-training database are HEAs with 3 – 20 at. % Al. This Al inclusion highlights that the strengthening effect from potent Al-refractory elements bonds has been inherently considered. The ML achieved a root-mean-square error (RMSE) of 9.2% for compressive $\varepsilon_f$ and 201 MPa for $\sigma_{YS}$ (Figure 2b,c), comparable to similar works [28]. The ML-identified important physical features are included in Table S2.

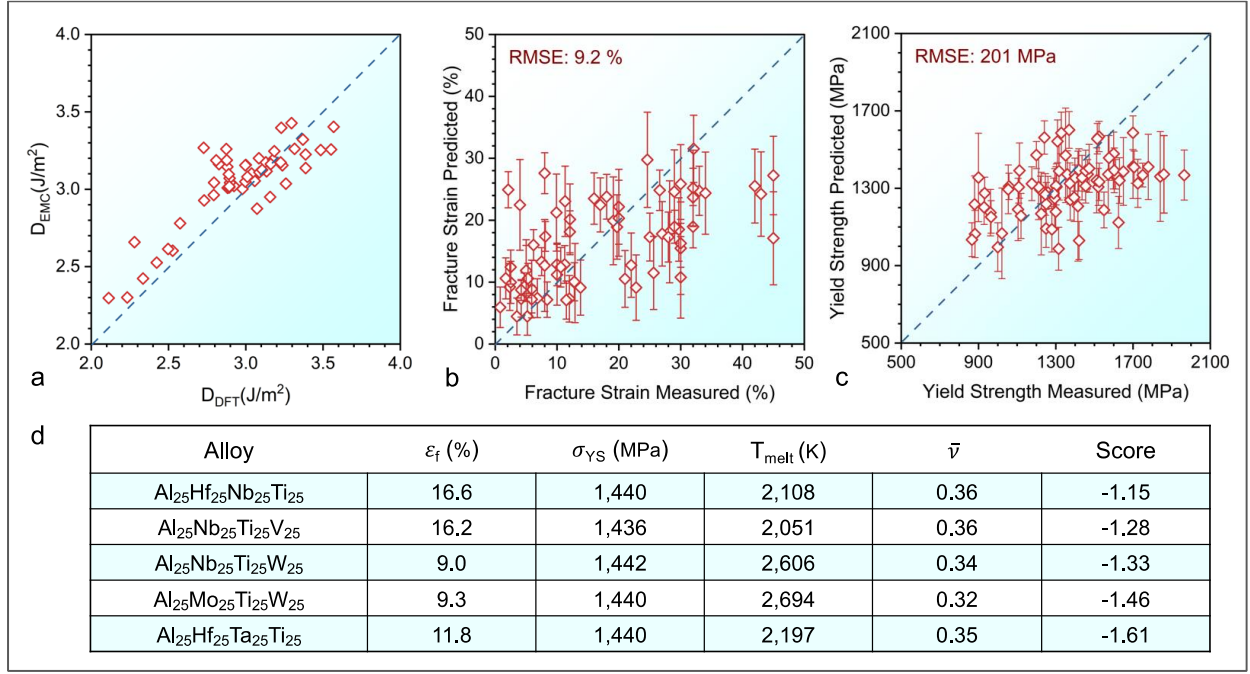

| Alloy | $\varepsilon_f$ (%) | $\sigma_{YS}$ (MPa) | $T_{melt}$ (K) | $\bar{v}$ | Score |
|---|---|---|---|---|---|
| Al$_{25}$Hf$_{25}$Nb$_{25}$Ti$_{25}$ | 16.6 | 1,440 | 2,108 | 0.36 | -1.15 |
| Al$_{25}$Nb$_{25}$Ti$_{25}$V$_{25}$ | 16.2 | 1,436 | 2,051 | 0.36 | -1.28 |
| Al$_{25}$Nb$_{25}$Ti$_{25}$W$_{25}$ | 9.0 | 1,442 | 2,606 | 0.34 | -1.33 |
| Al$_{25}$Mo$_{25}$Ti$_{25}$W$_{25}$ | 9.3 | 1,440 | 2,694 | 0.32 | -1.46 |
| Al$_{25}$Hf$_{25}$Ta$_{25}$Ti$_{25}$ | 11.8 | 1,440 | 2,197 | 0.35 | -1.61 |

Figure 2. (a) Comparison of D-Parameter values calculated using DFT and Effective Medium Calculation (EMC). The EMC method demonstrates a promising agreement with DFT results, yielding an absolute mean error of 5% and validating its potential for the efficient estimation of the D parameter. (b) A comparison between the compressive fracture strain, as predicted by ML and measured from experiments for BCC HEAs, is presented. A RMSE value of 9.2% is labeled. (c) Comparison between the compressive yield strengths, as predicted by ML and measured from experiments for BCC HEAs, is exhibited. A RMSE value of 201 MPa is labeled. Error bars for (b) and (c) are calculated, collecting all decision trees' predictions and computing the standard deviation for every alloy in the dataset. (d) The top five equimolar quaternary Al-RHEAs, which are predicted to form the B2 phase without other IM. Their predicted properties, including $\sigma_{YS}$ (yield strength), $\varepsilon_f$ (fracture strain), T$_{melt}$ (melting temperature), and $\bar{v}$ (Poisson's ratio), along with their respective scores (details in methods section), are also presented.

As depicted in Figure 1, the ICME model begins with 84 equimolar quaternary Al-RHEAs by combining Al with any three refractory elements. The top five systems forming the B2 phase without other undesired IM, and showing the best overall properties are listed in Figure 2d. (Details in the methods section; full list in Table S3). $Al_{25}Hf_{25}Nb_{25}Ti_{25}$ and $Al_{25}Nb_{25}Ti_{25}V_{25}$ are the leading candidates with similar predicted performance. Their high predicted $\varepsilon_f$ values suggest potential plasticity in the BCC phase, which may decrease with the B2-LRO formation. $Al_{25}Hf_{25}Nb_{25}Ti_{25}$, with Hf known to enhance grain-boundary adhesion [29], is chosen for initial experimental examination. Following composition adjustments for property optimization will be guided by experiments and MC studies.

**Mont Carlo Guided Experimental Optimization of Al-RHEAs**

*$Al_{25}Hf_{25}Nb_{25}Ti_{25}$*

The XRD and SEM characterizations of the initial composition, $Al_{25}Hf_{25}Nb_{25}Ti_{25}$ in the as-cast condition, show the single-B2-phase formation [Figure 3(a,b)]. This alloy exhibits high strength, with a compressive $\sigma_{YS}$ ~ 1.66 GPa, but lower $\varepsilon_f$ ~ 2.2 % (Figure 4a). The alloy exhibits strong B2-LRO, as indicated by the pronounced (100) superlattice XRD diffraction peak (Figure 3a). This extensive LRO could significantly hinder dislocation movement, resulting in a higher strength but decreased plasticity, compared to the values (Figure 2d) predicted assuming a disordered BCC phase forms.

Metropolis Monte Carlo (MC) simulations [17,30] (methods section) were conducted to study the LRO, the B2 order-disorder transition during solidification, and the atomic sublattice occupancy. $Al_{25}Hf_{25}Nb_{25}Ti_{25}$ shows pronounced B2-LRO (Figure 5a, LRO spans between 0 and 1, representing disordering and ordering) and atomic order parameters, $LRO_i$ (Figure 5c, $LRO_i$ spans between -1 and 1, with 0 and $\pm 1$ representing disordering and ordering), with the order-disorder transformation temperature, $T_{transform}$ ~ 2,000 K, slightly below the predicted melting temperature of $T_{melt}$ ~ 2,108 K. Given that phase transformation and evolution in as-cast HEAs continue below $T_{melt}$ due to high atomic kinetic energy (such rapid phase transition may cease ~ 0.8 $T_{melt}$ [21]), substantial B2-LRO can be developed in $Al_{25}Hf_{25}Nb_{25}Ti_{25}$ below the $T_{melt}$. Figure 5c shows that Al and (Hf, Ti) predominantly settle to different B2 sublattices, with Nb exhibiting little inclination. The site-occupancy tendency is also evident in Figure 5g, with Al-Ti and Al-Hf being the primary nearest-neighbor-pair around Al, Hf, and Ti. Self-pairs are less prevalent in all nearest-neighbor pairs. Such site occupancy can be attributed to the binary atomic interaction energies, $H_{ij}$, listed in Table S4, where Al-Hf and Al-Ti pairs exhibit the lowest $H_{ij}$, indicative of more stable bonding. These MC-discovered site-occupancy preferences have been validated in similar systems like $Ti_2AlHf$ [31] and AlNbTiV [32], by employing DFT, ML, MC simulations, and experiments [11].

The significant B2-LRO substantially reduces the plasticity from the predicted $\varepsilon_f$ value. Understanding the influence of each element on LRO is essential for the subsequent composition adjustment. Therefore, each element in $Al_{25}Hf_{25}Nb_{25}Ti_{25}$ was systematically reduced to 20 at.%, generating compositions of $Al_{20}(Hf_{26.7}Nb_{26.7}Ti_{26.6})$, $Hf_{20}(Al_{26.7}Nb_{26.7}Ti_{26.6})$, $Nb_{20}(Al_{26.7}Hf_{26.7}Ti_{26.6})$, and $Ti_{20}(Al_{26.7}Hf_{26.7}Nb_{26.6})$, referred to as $Al_{20}$, $Hf_{20}$, $Nb_{20}$, and $Ti_{20}$, respectively, in Figure 5b. $T_{transform}$ and the B2-LRO at specific reference temperatures (1,500 K as an example) are calculated. Decreasing the Al content lowers both $T_{transform}$ and the B2-LRO. Conversely, reducing Hf, Nb, or Ti raises $T_{transform}$ and LRO, likely due to the resultant increase in the Al content. This trend suggests that decreasing the Al content could be a strategy to lower B2-LRO and enhance ductility/plasticity. An additional heat treatment of $Al_{25}Hf_{25}Nb_{25}Ti_{25}$ also reveals a tendency of Laves-phase formation above 700°C (Figure S1). This phase transformation does not contradict the ICME phase prediction of a single B2 formation, which is predicted for the as-cast condition. The B2 region in the as-annealed samples has an EDS-determined composition of $Al_{20}Hf_{24}Nb_{29}Ti_{27}$ (Figure S1). Given its thermal stability and reduced Al content, this B2 composition is chosen for a more detailed study.

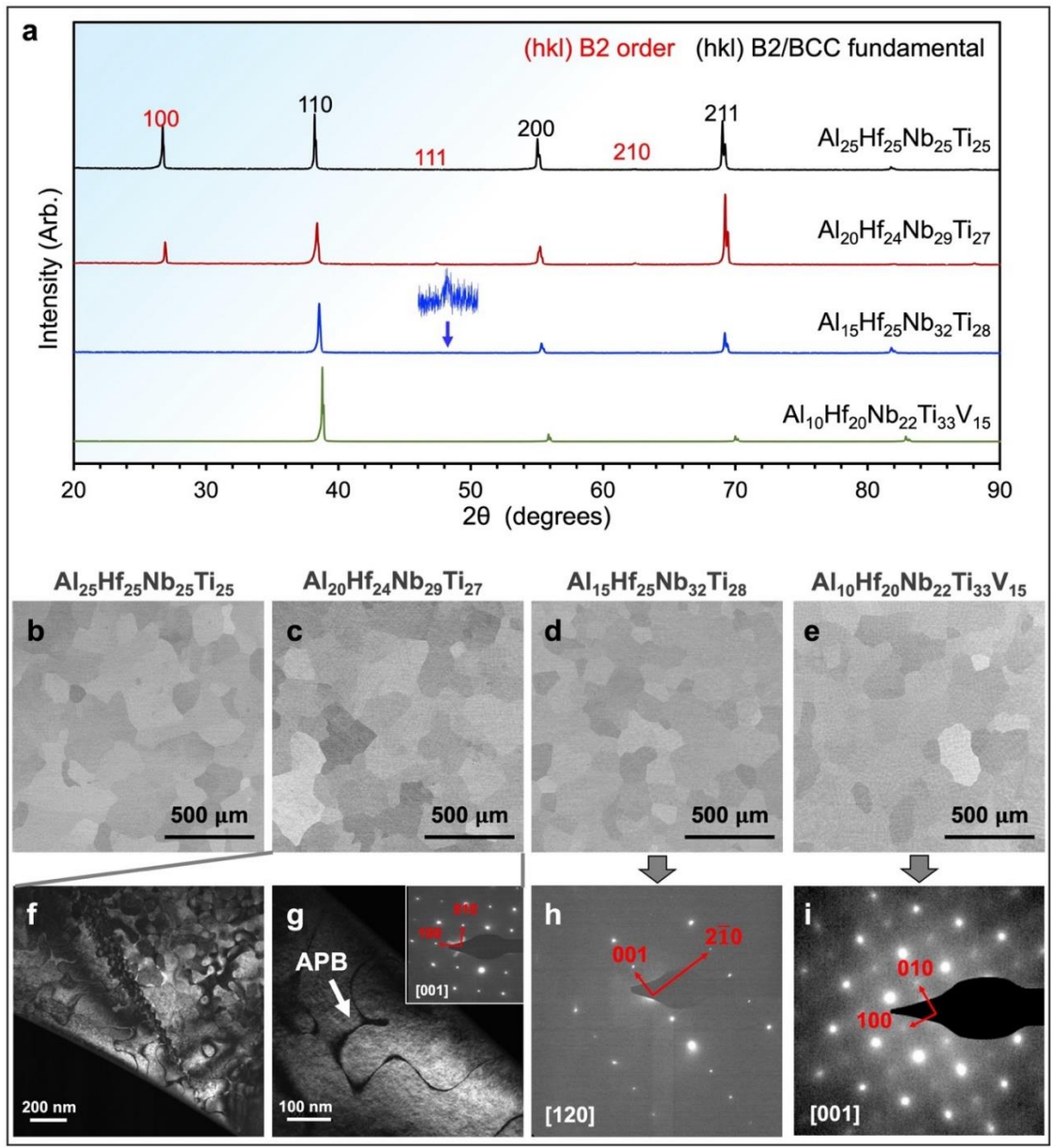

Figure 3. (a) XRD patterns for $Al_{25}Hf_{25}Nb_{25}Ti_{25}$, $Al_{20}Hf_{24}Nb_{29}Ti_{27}$, $Al_{15}Hf_{25}Nb_{32}Ti_{28}$, and $Al_{10}Hf_{20}Nb_{22}Ti_{33}V_{33}$ in the as-cast condition. Peaks corresponding to the disordered BCC phase or B2 phase are labeled with their (hkl) indices in black or red color. For $Al_{15}Hf_{25}Nb_{32}Ti_{28}$, the insets provide higher magnification views of the (111) superlattice diffraction peaks for the B2 phase; (b,c,d,e) SEM-Backscattered Electron (BSE) images for four alloys in the as-cast condition. B2-phase grains can be observed; (f,g) TEM-dark filed images for the B2 phase with an anti-phase boundary (APB) in the as-cast $Al_{20}Hf_{24}Nb_{29}Ti_{27}$. Inset of (g) is the selected area electron diffraction (SAED) showing B2-LRO; (h,i) SAED for the as-cast $Al_{15}Hf_{25}Nb_{32}Ti_{28}$ and $Al_{10}Hf_{20}Nb_{22}Ti_{33}V_{33}$ presenting B2-LRO.

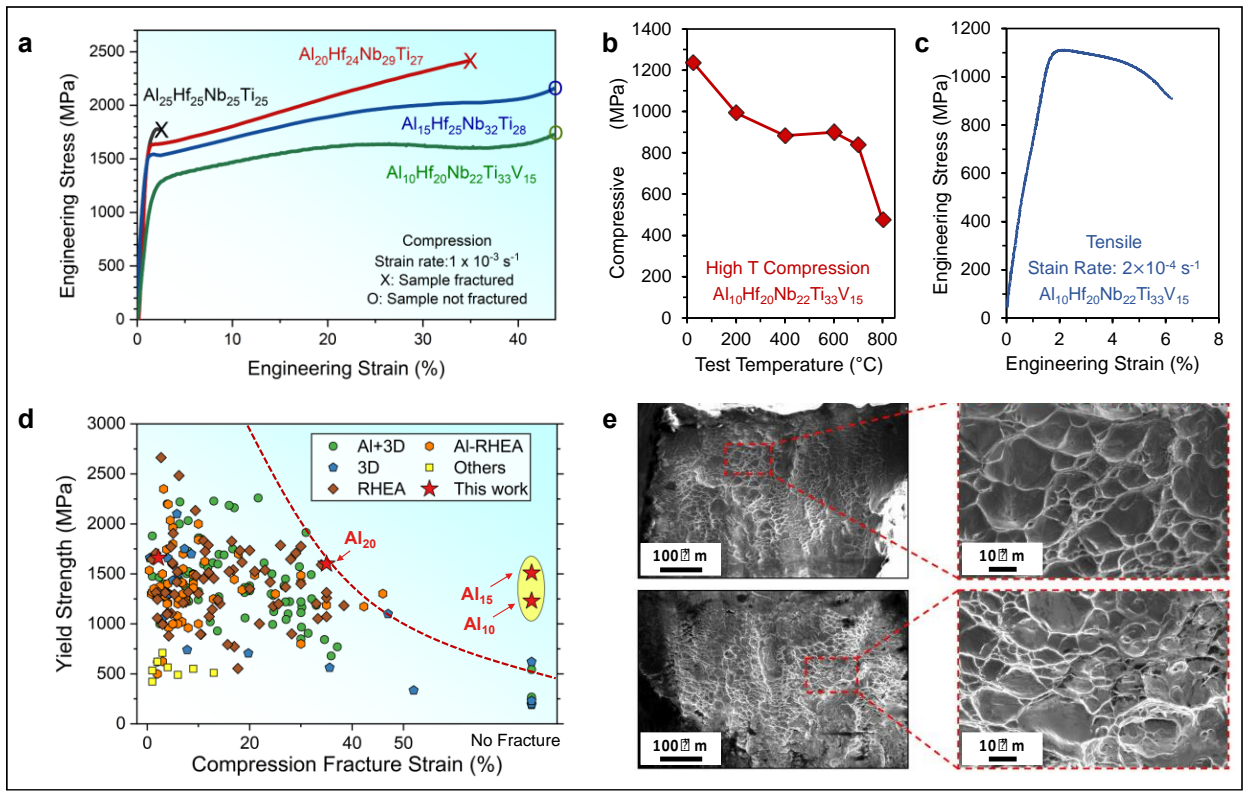

Figure 4. (a) Representative compression engineering stress-strain curves for the as-cast $Al_{25}Hf_{25}Nb_{25}Ti_{25}$, $Al_{20}Hf_{24}Nb_{29}Ti_{27}$, $Al_{15}Hf_{25}Nb_{32}Ti_{28}$, and $Al_{10}Hf_{20}Nb_{22}Ti_{33}V_{33}$. Tests terminated with sample fractured are labeled by crosses, while circles represent non-fractured samples; (b) High-temperature compression test showing the variation of the compressive yield strength against test temperature for $Al_{10}Hf_{20}Nb_{22}Ti_{33}V_{33}$; (c) Representative tensile engineering stress-strain curves for the as-cast $Al_{10}Hf_{20}Nb_{22}Ti_{33}V_{33}$; (d) Comparison of the compressive yield stress and fracture strain among HEAs in different categories; 3D represents HEAs with only 3d-transition metals. RHEAs denote HEAs with only refractory elements. Al+ represents the inclusion of Al. The two alloys in the present work with high toughness are highlighted with yellow oval. (e) SEM images showing the ductile fracture surfaces of the tensile test specimen in (c).

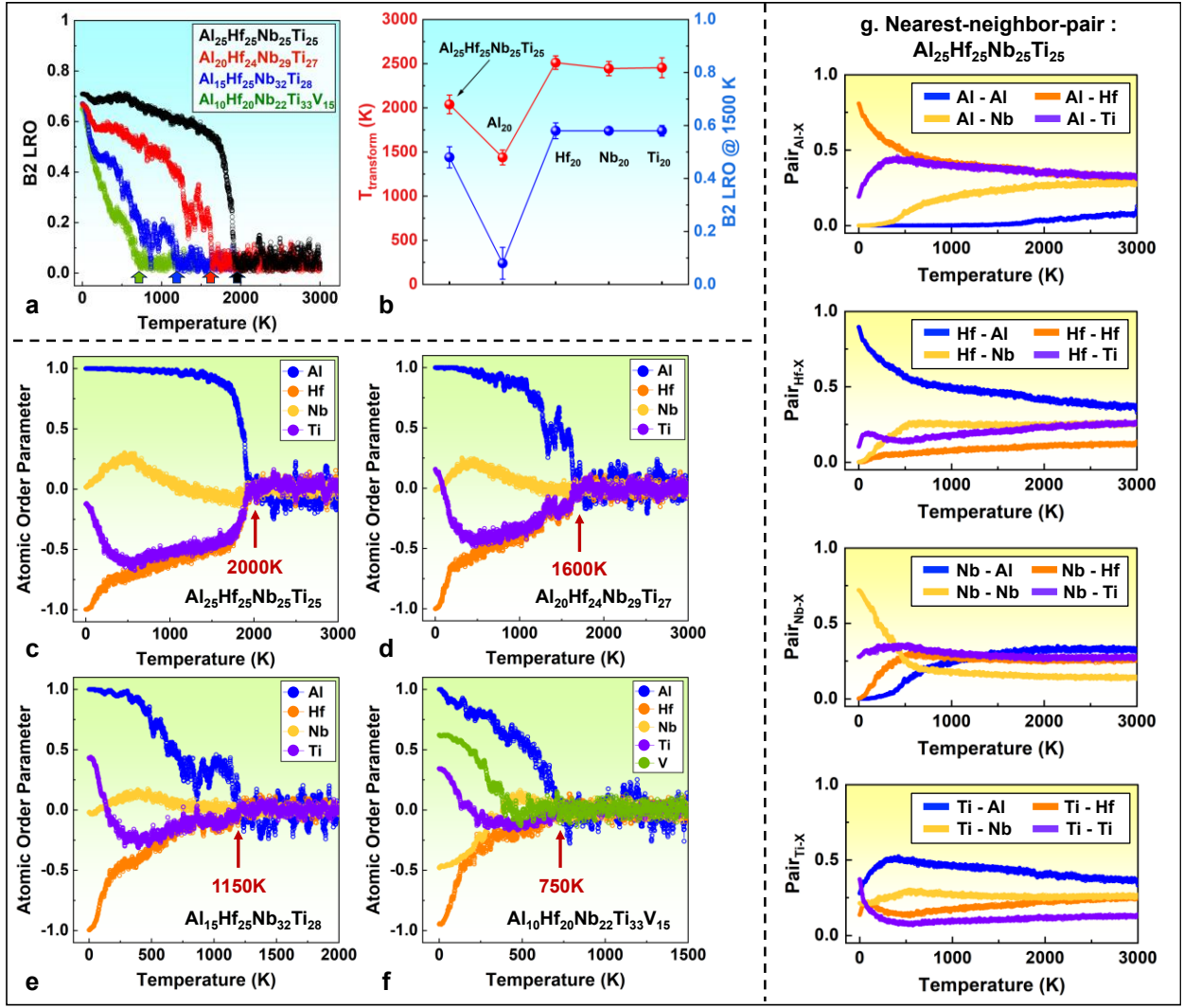

Figure 5. (a) Variation of long-range ordering (LRO) of the B2 phase under different temperatures in the four alloys: $Al_{25}Hf_{25}Nb_{25}Ti_{25}$, $Al_{20}Hf_{24}Nb_{29}Ti_{27}$, $Al_{15}Hf_{25}Nb_{32}Ti_{28}$, and $Al_{10}Hf_{20}Nb_{22}Ti_{33}V_{15}$ represented by black, red, blue, and green lines, respectively. LRO spans between 0 and 1, representing disordering and ordering. The points where B2-LRO disappears are labeled. (b) Variation of B2-ordering transformation temperature, $T_{transform}$, and B2-LRO at 1,500 K with different compositions. $Al_{20}$, $Hf_{20}$, $Nb_{20}$, and $Ti_{20}$ represent $Al_{20}(Hf_{26.7}Nb_{26.7}Ti_{26.6})$, $Hf_{20}(Al_{26.7}Nb_{26.7}Ti_{26.6})$, $Nb_{20}(Al_{26.7}Hf_{26.7}Ti_{26.6})$, and $Ti_{20}(Al_{26.7}Hf_{26.7}Nb_{26.6})$, respectively. Error

bars are from the standard deviations of five rounds of MC calculations. (c-f) Variation of atomic order parameters, $LRO_i$, under different temperatures for $Al_{25}Hf_{25}Nb_{25}Ti_{25}$, $Al_{20}Hf_{24}Nb_{29}Ti_{27}$, $Al_{15}Hf_{25}Nb_{32}Ti_{28}$, and $Al_{10}Hf_{20}Nb_{22}Ti_{33}V_{15}$. $LRO_i$ spans between -1 and 1, with 0 and ±1 denoting disordering and ordering. (g) Variation of nearest-neighbor-pair for different elements under various temperatures for $Al_{25}Hf_{25}Nb_{25}Ti_{25}$.

## *$Al_{20}Hf_{24}Nb_{29}Ti_{27}$*

The as-cast $Al_{20}Hf_{24}Nb_{29}Ti_{27}$ shows a single B2 phase from XRD and SEM characterizations in Figure 3(a,c). TEM investigations [Figure 3(f,g)] show the anti-phase boundaries, with the selected area electron diffraction (SAED, inset of Figure 3g) confirming the B2-LRO. Compression testing of $Al_{20}Hf_{24}Nb_{29}Ti_{27}$ (Figure 4a) exhibits a $\sigma_{YS}$ of 1.60 GPa, an ultimate strength of 2.41 GPa, and a notable $\varepsilon_f$ of 35 %. $Al_{20}Hf_{24}Nb_{29}Ti_{27}$ exhibits weakened B2-LRO than $Al_{25}Hf_{25}Nb_{25}Ti_{25}$, indicated by the less intense XRD-superlattice-diffractions and MC studies in Figure 5(a,d). The MC-computed $T_{transform}$ of ~ 1,600 K is close to 0.8 $T_{melt}$ where rapid phase transition still exists, allowing for B2-LRO development. Al and (Hf, Ti) continue their sublattice segregation (Figure 5d and S2). $Al_{20}Hf_{24}Nb_{29}Ti_{27}$ exhibits a significant $\varepsilon_f$ improvement, about sixteen times that of $Al_{25}Hf_{25}Nb_{25}Ti_{25}$, with nearly unchanged $\sigma_{YS}$. This improved plasticity could be attributed to the moderated B2-LRO, which facilitates plastic deformation through the glide of either <111> or a/2 <111> dislocations [33]. Interestingly, the LRO decrease does not significantly affect the overall strength. Next, a further Al reduction, as in $Al_{15}Hf_{25}Nb_{32}Ti_{28}$ with other elemental contents increased accordingly, may decrease the LRO and facilitate plasticity increase.

## *$Al_{15}Hf_{25}Nb_{32}Ti_{28}$*

The as-cast $Al_{15}Hf_{25}Nb_{32}Ti_{28}$ exhibits a single B2 phase (Figure 3d) with weak B2-LRO indicated by the magnified B2-(111) peak in the inset of Figure 3a. TEM-SAED (Figure 3h) further confirms the B2-LRO. The compression test (Figure 4a) exhibits a $\sigma_{YS}$ of 1.51 GPa. Notably, the samples endured up to 50%-height reduction without fracture before the test termination, demonstrating exceptional plasticity.

Compared to $Al_{25}Hf_{25}Nb_{25}Ti_{25}$ and $Al_{20}Hf_{24}Nb_{29}Ti_{27}$, $Al_{15}Hf_{25}Nb_{32}Ti_{28}$ experiences a marginal $\sigma_{YS}$ reduction by ~ 10 %, but a notable $\varepsilon_f$ increase. Concurrently, B2-LRO is further weakened to be nearly XRD-undetectable, suggesting that LRO exerts a more pronounced effect on plasticity rather than strength. The MC studies in Figure 5(a,e) show a $T_{transform}$ ~ 1,150 K, around half of the predicted $T_{melt}$ ~ 2,208 K. The limited atomic mobility and diffusion at $T_{transform}$ leads to only minimal LRO establishment.

### $Al_{10}Hf_{20}Nb_{22}Ti_{33}V_{15}$

We demonstrated that an Al reduction in AlHfNbTi system reduces B2-LRO, improving plasticity and slightly reducing strength. Based on this feature, we further explored integrating the top two ICME-identified quaternary alloy systems, AlHfNbTi and AlNbTiV (Figure 2d). By partially substituting Hf with V, we anticipate an alloy with lower cost/density, while retaining excellent plasticity and grain-boundary adhesion [29] from Hf. The Al content was further reduced to enhance ductility/plasticity, leading to the creation of $Al_{10}Hf_{20}Nb_{22}Ti_{33}V_{15}$, a stable single B2-phase alloy with compression plasticity and tensile ductility.

The XRD and SEM characterizations [Figure 3(a,e)] reveal a single phase and without B2-superlattice diffraction in the XRD pattern. Yet, B2-LRO is observable in the TEM-SEAD (Figure 3i). The compression test (Figure 4a) yields a $\sigma_{YS}$ of 1.23 GPa, with no fracture up to 50%-height reduction, aligning with the reduced B2-LRO leading to lower strength but more plasticity. High-temperature compression test (Figure 4b) maintains $\sigma_{YS}$ above 800 MPa at 700°C, with a rapid strength decline around 800°C. The tensile test (Figure 4c) exhibits a $\sigma_{YS}$ of 1.10 GPa and $\varepsilon_f$ of 6.3 %, with tensile fracture surfaces (Figure 4e) further substantiating the ductility.

The MC results [Figure 5(a,f)] show a $T_{transform}$ ~ 750 K, markedly below the predicted $T_{melt}$ ~ 2,058 K. Nonetheless, $Al_{10}Hf_{20}Nb_{22}Ti_{33}V_{15}$ manifests a faint B2 ordering, likely attributable to the restrained yet present atomic mobility around $T_{transform}$. Al and (Hf, Ti) segregate to different sublattices, while the Nb and V exhibit weak superlattice occupancy tendencies when cooling to slightly below $T_{transform}$ (Figure 5f). A further temperature decline below 500 K leads to pronounced elemental segregation driven by the nominal impact of entropy. However, this low-temperature region is less important as achieving the MC-predicted equilibrium state is nearly impossible given the exceptionally sluggish atomic diffusion.

**Comparative Study of Al-RHEAs Mechanical Properties**

Around 200 HEAs are compared for compressive $\sigma_{YS}$ and $\varepsilon_f$ in Figure 4d, with the red line highlighting the trend of increased strength with decreased plasticity. Most Al-RHEAs exhibit limited plasticity, potentially due to: alloy's inherent brittleness [34], as elaborated in the next section; strong Al-induced B2-LRO; and brittle IM formation [11,13]. However, Al-RHEAs designed in the present work diverge from this norm, achieving both high strength and plasticity/ductility. These ICME-designed Al-RHEAs exhibit inherent ductility and avoid brittle IM formation. Subsequent experimental-MC hybrid optimizations are applied to adjust the LRO, thereby enhancing the plasticity/ductility of the alloy without compromising its strength.

**VEC Valley of Metal Brittleness and DFT Analysis**

During the ICME ML-training process, we identified an intriguing correlation between VEC [18] and HEA plasticity. Figure 6a shows a "VEC valley" of brittleness between 5.5 and 6.2 in graph plotting compressive $\sigma_{YS}$ against VEC. This valley is not attributed to data deficiency, as we specifically targeted HEAs in this VEC range. Previous ML-studies [35,36] have recognized the strong VEC-plasticity/ductility correlations, and the D parameter, known to indicate plasticity/ductility, also correlates with VEC [25,26]. Notably, our Al-RHEAs consistently locate in the lower-VEC-regions, outside the valley, where high plasticity/ductility HEAs are typically found.

Empirical studies [18] suggest a "spectrum" of the VEC-HEA phase formation, where VEC < 6 favors FCC formation, 6 < VEC < 7.8 leads to a mixed FCC-BCC phase, and VEC > 7.8 to single BCC phase. The range 6.88 < VEC < 7.84 tends to form brittle topologically close-packed (TCP) phases (e.g., Laves, Sigma) [37]. The "VEC valley" resides near the higher BCC and mixed FCC-BCC regions, close to the brittle TCP region. To the right of the valley, HEAs form typically ductile FCC phase, while to the left, ductile BCC phases. A decreasing VEC lowers the Fermi level in the band structure, leading to an earlier occurrence of the critical strain for shear instability and thus inherently enhancing the plasticity/ductility [34]. Therefore, the "VEC valley" only occupies the higher end of the BCC region.

Beyond these reasons, a more in-depth, quantitative analysis is critical for deciphering the fundamental mechanisms that underpin this phenomenon, and DFT was applied to provide more insight. Prior studies [38,39] suggest that the Fermi level density of states, $D(E_F)$, correlates with plasticity/ductility due to the strength of metallic bonding relative to ionic and covalent bonds. Other work has linked high plasticity/ductility to incipient shear instabilities governed by the proximity of the Fermi level to topologically unstable energy-band crossings [40,41]. Electronic-structure effects may thus help explain the nonmonotonic variation of fracture strain with respect to VEC. Figure 6b presents $D(E)$ for a representative refractory metal, BCC Mo, with VEC = 6 at the center of the experimental region of interest. The pronounced pseudogap around the Fermi level at $E_F = 0$ is caused by the $E_g - T_{2g}$ splitting at the $\Gamma$ $k$-point with the Mo $d$-band half-filled in its standard electron configuration, $5s^1 4d^5$. In a rigid band model [42], we may assume that $D(E)$ for the alloys under study qualitatively resembles that of Mo, with the chief difference (other than the possible magnetism) being the placement of the Fermi level, which grows monotonically with VEC. The "VEC valley" for brittleness (Figure 6b) roughly coincides with the region of relatively low $D(E_F)$.

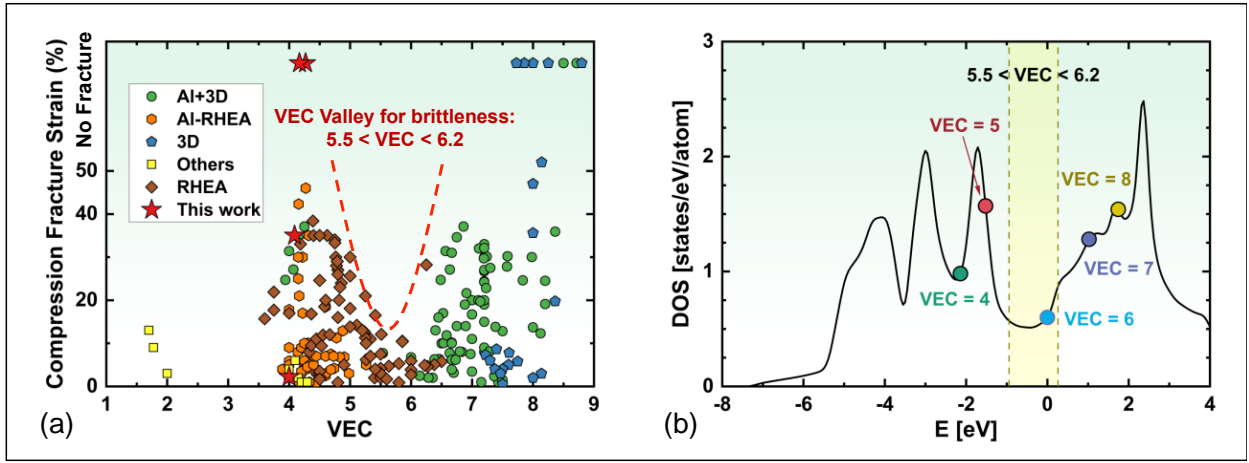

Figure 6. (a) Comparison of the compressive fracture strain and VEC among HEAs in different categories. In both figures, 3D represents HEAs with only 3d-transition metals. RHEAs denote HEAs with only refractory elements. Al+ represents the inclusion of Al. The four alloys in the current work, in both AC and 1,100°C-annealed conditions, are highlighted by the magenta points. (b) The density of states (DOS) of Mo showing the variation of $D(E_F)$ with respect to the valence electron count (VEC) in a rigid band model. Plotting symbols are placed at energies corresponding to VECs of the periodic table columns 4 - 8 (Ti-Fe columns). The Fermi energy for VEC = 6 is set to zero. The brittle region (5.5 < VEC < 6.2) is shaded.

## 3. Discussion

HEAs offer a remarkable latitude in designing alloys with multiple promising properties. This potential, however, brings the complex challenge of systematically exploring the HEA compositional space for optimal compositions suitable for various applications. The past decade has witnessed a surge in the HEA-phase prediction research [23], and recognized ML and CALPAHD as ideal computational tools for high-throughput computation. Concurrently, many properties-prediction models emerged with high accuracies [35,36,43], and paved the way for the inception of all-encompassing ICME HEA design models. The discovery of the strong, yet ductile Al-RHEAs epitomizes a refined template model, and the ICME framework and sub-models can be customized accordingly to divergent design objectives. However, while ICME effectively identifies promising compositional zones, expecting it to determine the optimal composition single-handedly is still ambitious. As shown with the AlHfNbTi(V) system, finetuning the composition necessitates deeper theoretical analysis, such as MC and DFT, complemented by experimental validation and refinement.

To review the AlHfNbTi(V) system, four B2 Al-RHEAs were designed with plasticity greatly determined by B2-LR and adjustable through the Al content. MC studies suggest that the LRO may also be modulated through specific thermal treatments, tailoring ductility for specific applications. Notably, AlHfNbTi(V) alloys demonstrate toughness surpassing most extant HEAs, with the Al incorporation effectively reducing cost and density while enhancing oxidation resistance, positioning this system promising for continued optimization and research.

Regarding the fundamental mechanism of the VEC-plasticity correlation, first-principles calculations have elucidated an electronic origin of the "VEC valley." The more in-depth investigation is imperative, propelling the field toward an enriched understanding of electronic structure effects on alloy ductility. Concurrently, strategically avoiding the "VEC valley" is advised in designing future ductile alloys.

Moving ahead, we envision this novel alloy-design strategy as a blueprint for designing alloys with structural applications. We also aim to deepen understanding of elements' role in atomic ordering and mechanical properties in high-entropy alloys and explore VEC-centric compositional tuning to overcome brittleness, thereby perfecting the approach and enriching the theoretical framework.

## 4. Methods

**Effective Medium Calculations of D Parameter and ML Mechanical Properties Prediction Model**

The D parameter quantifies the ratio between the surface energy associated with the cleavage fracture ($\gamma_{sf}$) and the energy associated with the unstable stacking fault ($\gamma_{usf}$) [25]. A distinct correlation has been observed between the compression fracture strain and the DFT-calculated D parameter for BCC HEAs [25] because it has been postulated that the ductility of BCC alloys is influenced by the interplay between the dislocation emission and cleavage-fracture propagation in the vicinity of a crack tip [26]. Despite its importance, the D parameter computed by DFT is computationally expensive and not suitable for high-throughput simulations. Alternatively, in our effective medium method [24], the D parameter is calculated by considering the values of surface energies and unstable stacking fault energies along the same atomic planes and dislocation directions for each constituent element with the equation and details listed in Table S1.

The ML-mechanical-properties prediction database includes 72 data entries, with the initial features listed in Table S1 and processed by Feature Engineering [22]. A Genetic Algorithm [44] was employed for the optimal feature selection. Multiple ML regression algorithms, such as the Support Vector Regressor, Random Forest Regressor (RFR), K-Nearest Neighbor, and Linear Regression algorithms, were tested, and RFR was adopted for its overall superior performance.

**ICME Alloy System Selection:**

Out of the 84 equimolar quaternary Al-RHEAs, sixteen Al-RHEAs systems are screened out by the ML prediction of forming the B2 phase without other IMs. These systems are then predicted for the compressive $\sigma_{YS}$, $\varepsilon_f$, $T_{melt}$ (Equation 7), and $\bar{v}$ (Equation in Table S1). Three heavier Al-RHEAs, with ρ exceeding 10 g/cc, are excluded due to their significantly higher density, compared to Ni- and Co-based superalloys. The predicted property values for each system are summarized in Table S3. Each alloy is scored using:

$$\text{score} = \sum_{\text{all } i} \frac{p_i - p_{i-\text{best}}}{p_{i-\text{range}}} \qquad \text{Equation 1}$$

where $p_i$ represents the predicted value of the i-th property for the alloy, $p_{i-\text{best}}$ and $p_{i-\text{range}}$ are, respectively, the best value (i.e., the highest $\varepsilon_f$, $\sigma_{YS}$, $T_{melt}$, and $\bar{v}$) and the range for the i-th property among all the candidates. A higher score represents a better overall performance.

**Experiment methods**

High-purity raw elements (> 99.99 weight percent, wt.%) were arc-melted in a water-cooled copper crucible using a 400A current. Each melt lasted 30 seconds, with a total of five melts performed, and the ingots were flipped between each melt. The ingots were later polished for X-ray diffraction (XRD) and Scanning Electron Microscope (SEM) analysis. Parts of the ingots were also suction-cast into copper molds to produce: (1) a rod-shaped sample (3 mm diameter, 10 mm height) for compression testing and Transmission Electron Microscope (TEM) sample preparation; (2) a rectangular bar (2 mm x 5 mm x 10 mm) later machined into a dog-bone shape for tensile testing.

The XRD/SEM samples were polished down to a grit size of 1 μm with diamond suspension and finished with a 0.06-μm colloidal silica suspension. The TEM samples were sliced from the suction-cast rod with a thickness of 500 μm. These thin slices were further polished to a thickness below 100 μm, followed by electron polishing to create TEM-transparent regions for imaging.

The XRD measurements were conducted on the polished sample by a PANalytical Empyrean diffractometer with Cu Kα radiation and a scanning rate of approximately 0.15 degrees/s. The SEM characterization was conducted on an FEI Quanta 650 equipped with Energy Dispersive Spectroscopy (EDS). The TEM analysis was performed, using an FEI Titan Transmission Electron Microscope.

Cylindrical samples with 3 mm in diameter and 6 mm in gauge length were used for compression testing, and dog-bone specimens with a cross-sectional area of less than 1 mm × 1 mm and a gauge length of ~ 5 mm were used for tensile testing. Compression tests were performed on a computer-controlled, uniaxial mechanical testing system (an MTS servo-hydraulic load-frame) with an initial strain rate of $1 \times 10^{-3}$ s$^{-1}$. The load and displacement values of the testing system were recorded without specimens, and these displacement values were subtracted from the measured displacement values in compression tests at their corresponding load. This method excludes the elastic responses of the load frame, load cell, and sample grips. Then the deformation of the specimens could be calculated. Tensile tests with the strain rate of $2 \times 10^{-4}$ s$^{-1}$ were conducted, using an Instron-Electroplus E1000 electrodynamic test instrument, which can accurately control the small loads for miniature specimens. The load and displacement values of the system were measured by loading a dummy sample with a known elastic modulus. After subtracting the elastic response of the dummy sample, the calculated load and displacement values of the testing system could be subtracted from the measurements.

**Density functional theory**

The first-principles calculations were performed within the electronic density functional theory [45,46], using the Vienna Ab initio Simulation Package (VASP) code [47,48] in the Perdew-Burke-Ernzerhof (PBE) generalized gradient approximation [49]. The lattice parameters were relaxed with a high *k*-point density and default plane-wave energy cutoff. The Mo density of states (DOS) was calculated, using tetrahedron integration with a subsequent 0.1 eV Gaussian smearing, and we have shifted the energy scale to place the Fermi energy of Mo at zero. Fermi energies for VEC = 4 – 8 were calculated from the integrated density of states.

**Monte Carlo Simulation**

The Metropolis MC simulation [17,30] incorporates the nearest-neighbor interaction paradigm with the nearest-neighbor interaction energy, $v_{ij}$, given by:

$$v_{ij} = \frac{H_{ij}}{z} \qquad \text{Equation 2}$$

Here, $H_{ij}$ is the binary energy of formation [50]. The parameter, z, is the number of the nearest-neighbor bonds per atom and has a value of 4 in the BCC structure [17]. $v_{ij}$ values used in this study are tabulated in Table S4.

The system commences with a randomized structural configuration, mirroring the disordered alloy's atomic disposition. Within every MC iteration, a random change to the current configuration is proposed. In this study, this random alteration is the positional exchange of two atoms. The energies of the old and new configurations are compared. If the new configuration has lower energy, it is accepted. If it has higher energy, it is accepted with a Boltzmann probability, P:

$$P = e^{\frac{-\Delta H}{k_B T}} \qquad \text{Equation 3}$$

where $\Delta H$ is the energy difference between the two configurations, $k_B$ is the Boltzmann constant, and T is the temperature. The MC simulation in this study uses a supercell configuration of 12 × 12 × 24 comprising 3,456 atoms and executes over $10^7$ MC iterations.

The BCC/B2 lattice comprises two distinct sublattices, α and β. In a disordered BCC phase, elements are uniformly distributed between α and β, while in an ordered B2 matrix, they show sublattice preference. The order parameter for atom, i, denoted as $LRO_i$, is given by:

$$LRO_i = \frac{x_{i-\alpha} - x_{i-\beta}}{x_{i-\alpha} + x_{i-\beta}} \qquad \text{Equation 4}$$

Here, $x_{i-\alpha}$ and $x_{i-\beta}$ represent the occupancy of the element, i, on the α and β sublattices, respectively. The $LRO_i$ value spans from -1 to 1, where 0 denotes a lack of sublattice preference. Conversely, values of 1 or -1 indicate an exclusive occupancy of a particular sublattice, representing the highest degree of atomic ordering.

The alloy's LRO is defined as:

$$LRO = \sqrt{\sum_{\text{all i}} c_i \, LRO_i^2} \qquad \text{Equation 5}$$

where $c_i$ denotes the atomic percentage of each kind of atom, i.e., LRO spans between 0 (disordered) and 1 (ordered).

The ratio of the nearest-neighbor pairs, $Pair_{i-j}$, is defined as:

$$Pair_{i-j} = \frac{n_{i-j}}{\sum_{\text{all j}} n_{i-j}} \qquad \text{Equation 6}$$

within this context, $n_{i-j}$ represents the count of an atom pair, i–j, in the MC supercell. The $Pair_{i-j}$ quantifies the probability of atom, j, being among the nearest neighbors of atom, i.e., $Pair_{i-j}$ generally remains steady at elevated temperatures with a disordered BCC phase and is only determined by the atomic percentages of atoms, i and j, in the alloy. As the temperature decreases, certain nearest-neighbor pairs gain preference, leading to a deviation of the $Pair_{i-j}$ distribution from a random arrangement. This shift underscores the dynamic nature of atomic interactions and the emerging order within the system.

**Alloy Melting Temperature and Density Prediction Method**

$T_{melt}$ is predicted by the following method [21]:

$$T_{melt} = \frac{\sum_{i \neq j} T_{i-j} \times c_i \times c_j}{\sum_{i \neq j} c_i \times c_j} \qquad \text{Equation 7}$$

Here, $T_{i-j}$ is the binary liquidus temperature of the element pair, i-j, on the binary-phase diagram with a relative atomic ratio of $c_i/c_j$. The method of extracting binary liquidus temperatures is described in detail in the previous work [21].

The density, $\rho$, is predicted by the rule of mixtures:

$$\frac{1}{\rho} = \sum_i \frac{w_i}{\rho_i} \qquad \text{Equation 8}$$

Here, $w_i$ and $\rho_i$ are the weight percentage and density of the i-th element.

## 5. Acknowledgments

MW is supported by the Department of Energy under Grant No. DE-SC0014506 for the study of the electronic density of states. MW acknowledges a discussion with Amit Samanta on the link between the Fermi level density of states and ductility. JQ, DH, and JP are supported by the Office of Naval Research under Grant No. N00014-23-1-2441. XF and PKL very much appreciate the support from (1) the National Science Foundation (DMR – 1611180, 1809640, and 2226508) and (2) the US Army Research Office (W911NF-13–1-0438 and W911NF-19–2-0049).

## 6. Data Availability

The data supporting this work are available from the authors upon reasonable request.

# Supplementary Materials

Jie Qi, Xuesong Fan, Diego Ibarra Hoyos, Michael Widom, Peter K. Liaw, and S. Joseph Poon, Integrated Design of Aluminum-Containing High-entropy Refractory B2 Alloys with Synergy of High Strength and Ductility, 2023

Table S1: Definition of physics-based features used in the Machine-Learning Mechanical-Properties model.

| Features | Details |
|---|---|
| $\frac{\gamma_{sf}}{\gamma_{usf}}$ **(D Parameter)** $\equiv \frac{\sum_i c_i * V_i * \frac{\gamma_{sf\_i}}{\gamma_{usf\_i}}}{\sum_i c_i * V_i}$ [1] | $c_i$: Atomic percentage for the $i_{th}$ element in an n-component system. (Definitions of n and $c_i$ are the same elsewhere.) $V_i$: Volume fraction for the $i_{th}$ element $\gamma_{sf\_i}$: Surface energy for the $i_{th}$ element [2] $\gamma_{usf\_i}$: Unstable stacking fault energy for the $i_{th}$ element [3–10]. |
| **μ (Shear Modulus)** $\equiv \sum_{i=1}^{n} \frac{1}{\sum_{i=1}^{n} c_i * \frac{1}{\mu_i}}$ [11,12] | $\mu_i$: Shear modulus of the $i_{th}$ element |
| **E (Young's Modulus)** $\equiv \sum_{i=1}^{n} \frac{1}{\sum_{i=1}^{n} c_i * \frac{1}{E_i}}$ [11,12] | $E_i$: Young's modulus of the $i_{th}$ element |
| **K (Bulk Modulus)** $\equiv \sum_{i=1}^{n} \frac{1}{\sum_{i=1}^{n} c_i * \frac{1}{K_i}}$ [11,12] | $K_i$: Bulks modulus of the $i_{th}$ element |
| **$\bar{\nu}$ (Poisson's Ratio)** $\equiv \sum_{i=1}^{n} c_i * \nu_i$ [11,12] | $\nu_i$: Poisson's ratio of the $i_{th}$ element |
| $VEC \equiv \sum_{i=1}^{n} c_i * VEC_i$ [13] | $VEC_i$: Valence Electron Count of the $i_{th}$ element |
| $\Delta H_{mix} \equiv \sum_{i=1}^{n} 4 \Delta H_{i,j}^{mix} c_i c_j$ [13] | $\Delta H_{i,j}^{mix}$: The binary mixing enthalpy obtained from the Miedema's model [14] of the i-j element pair. |
| $\Delta S_{mix} = -R \sum_{i=1}^{N} c_i \ln(c_i)$ [13] | R: The gas constant. |
| $\Omega \equiv T_m \Delta S_{mix}/|\Delta H_{mix}|$ [13] | $T_m$: Alloy melting temperature. |
| $\varepsilon_{ave}^{4/3}$ (Average atomic size misfit) $\equiv$ $\sum_{ij} \left\| \frac{1}{a} \frac{da}{dx_i^j} \right\| c_i c_j$ [15] | a = Average lattice constant. $\frac{da}{dx_i^j}$ = Variation of the lattice constant when a small amount of element, i, in the average matrix of the alloys is replaced by an element, j. |
| $\delta_\chi$ (Electronegativity mismatch) $\equiv$ $\frac{\sum_{i=1}^{n} \sum_{j=1, i \neq j}^{n} c_i c_j * \frac{|\chi_i - \chi|}{\chi}}{\sum_{i=1}^{n} \sum_{j=1, i \neq j}^{n} c_i c_j}$ [13] | $\chi_i$: Electronegativity of the $i_{th}$ element. $\chi = \sum_{i=1}^{n} c_i \chi_i$: Average electronegativity. |
| Pressure field $\equiv \frac{\mu(1+\nu)}{3\pi(1-\nu)}$ [16] | |
| $E_2/E_0$ (Geometric Strain) $\equiv$ $\sum_{j \geq i}^{n} \frac{c_i * c_j * |r_i + r_j - 2r|^2}{(2r)^2}$ [13] | $r = \sum_{i=1}^{n} c_i r_i$: Average atomic radius. |

Table S2: An overview of the most relevant features selected by the Genetic Algorithm (GA) for Fracture Strain and Yield Strength ML models for BCC HEAs.

| Fracture strain model | Yield strength model |
|---|---|
| $\ln(\varepsilon_{ave}^{4/3}) / (\frac{\gamma_{sf}}{\gamma_{usf}})^2$ | $\frac{\gamma_{sf}}{\gamma_{usf}} \cdot \ln(K)$ |
| $\ln(E_2/E_0) - (\varepsilon_{ave}^{4/3})^2$ | $\ln(\varepsilon_{ave}^{4/3}) - (\frac{\gamma_{sf}}{\gamma_{usf}})^2$ |
| $\ln(\delta_\chi)/\varepsilon_{ave}^{4/3}$ | $(E_2/E_0)^2 - \frac{1}{VEC}$ |

Table S3: This table lists thirteen equimolar quaternary Al-RHEAs, which are predicted to form the B2 phase without other IM phases. Their predicted properties, including $\sigma_{YS}$ (yield strength), $\varepsilon_f$ (fracture strain), T$_{melt}$ (melting temperature), and $\bar{\nu}$ (Poisson's ratio), along with their respective scores, are also presented.

| Alloy | $\varepsilon_f$ (%) | $\sigma_{YS}$ (MPa) | T$_{melt}$ (K) | Poisson's ratio | Score |
|---|---|---|---|---|---|
| Al$_{25}$Hf$_{25}$Nb$_{25}$Ti$_{25}$ | 16.6 | 1,440 | 2,108 | 0.36 | -1.15 |
| Al$_{25}$Nb$_{25}$Ti$_{25}$V$_{25}$ | 16.2 | 1,436 | 2,051 | 0.36 | -1.28 |
| Al$_{25}$Nb$_{25}$Ti$_{25}$W$_{25}$ | 9.0 | 1,442 | 2,606 | 0.34 | -1.33 |
| Al$_{25}$Mo$_{25}$Ti$_{25}$W$_{25}$ | 9.3 | 1,440 | 2,694 | 0.32 | -1.46 |
| Al$_{25}$Hf$_{25}$Ta$_{25}$Ti$_{25}$ | 11.8 | 1,440 | 2,197 | 0.35 | -1.61 |
| Al$_{25}$Mo$_{25}$Nb$_{25}$Ti$_{25}$ | 6.7 | 1,441 | 2,282 | 0.35 | -1.92 |
| Al$_{25}$Ta$_{25}$Ti$_{25}$V$_{25}$ | 8.3 | 1,436 | 2,147 | 0.35 | -2.00 |
| Al$_{25}$Hf$_{25}$Ti$_{25}$Zr$_{25}$ | 7.1 | 1,496 | 1,983 | 0.35 | -1.98 |
| Al$_{25}$Mo$_{25}$Ta$_{25}$Ti$_{25}$ | 4.9 | 1,441 | 2,329 | 0.33 | -2.17 |
| Al$_{25}$Mo$_{25}$Ti$_{25}$V$_{25}$ | 6.0 | 1,441 | 2,171 | 0.34 | -2.22 |
| Al$_{25}$Ti$_{25}$V$_{25}$W$_{25}$ | 6.0 | 1,322 | 2,504 | 0.33 | -2.51 |
| Al$_{25}$Cr$_{25}$Ti$_{25}$W$_{25}$ | 9.9 | 1,341 | 2,407 | 0.29 | -2.66 |
| Al$_{25}$Cr$_{25}$Mo$_{25}$Ti$_{25}$ | 6.3 | 1,385 | 2,075 | 0.27 | -3.39 |

Table S4: Nearest neighbor enthalpy of formation, $H_{ij}$ (meV/atom) [17].

| Elements | Al | Hf | Nb | Ti | V |
|---|---|---|---|---|---|
| Al | 0 | -444 | -288 | -428 | -282 |
| Hf | -444 | 0 | 23 | -10 | 7 |
| Nb | -288 | 23 | 0 | 11 | -56 |
| Ti | -428 | -10 | 11 | 0 | 37 |
| V | -282 | 7 | -56 | 37 | 0 |

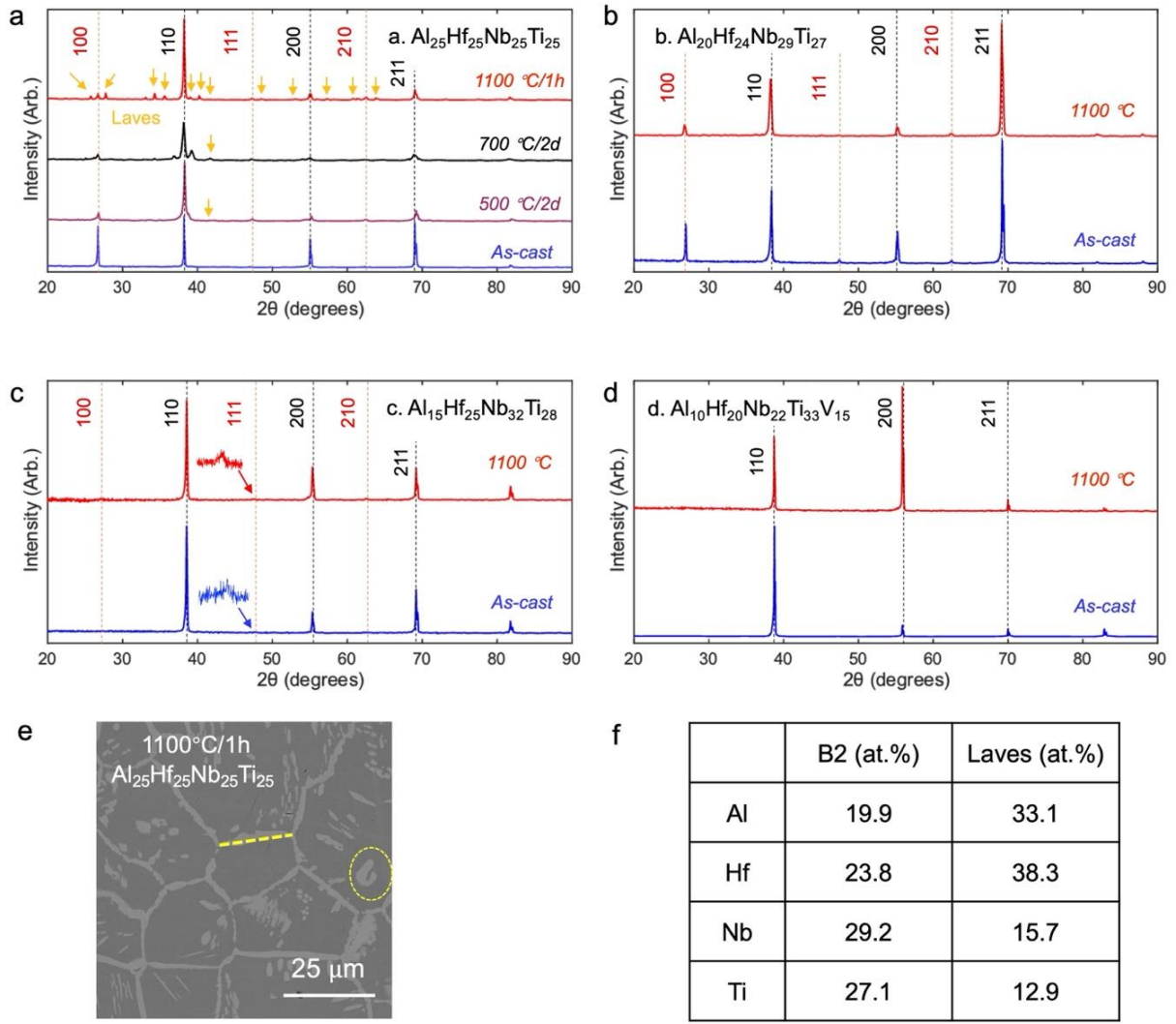

Figure S1. (a) X-ray diffraction (XRD) patterns for the $Al_{25}Hf_{25}Nb_{25}Ti_{25}$ in as-cast, 500 °C/2 days, 700 °C/2 days, and 1,100°C/1 hour-annealed conditions. (hkl) in red color are B2 super-diffraction peaks. The formation of a Laves phase is observed at temperatures above 700 °C, with a potential onset at 500 °C; (b-d) XRD patterns for the $Al_{20}Hf_{24}Nb_{29}Ti_{27}$, $Al_{15}Hf_{25}Nb_{32}Ti_{28}$, and $Al_{10}Hf_{20}Nb_{22}Ti_{33}V_{15}$ in as-cast and 1,100°C/1 hour-annealed conditions. Insets in (c) present an enlarged view of the B2-111 peaks; (e) Scanning Electron Microscope-Backscattered Electron (SEM-BSE) images of $Al_{25}Hf_{25}Nb_{25}Ti_{25}$ in the 1,100°C/1 hour-annealed condition. Areas indicative of the Laves phase, displaying lighter contrast, are demarcated with yellow dashed lines;

(f) Energy-Dispersive X-ray Spectroscopy (EDS) determined compositions of the B2 matrix and Laves phase in (e).

Table S4. (EDS) determined overall compositions (at. %) of the four alloys studied in the present work.

|     | $Al_{25}Hf_{25}Nb_{25}Ti_{25}$ | $Al_{20}Hf_{24}Nb_{29}Ti_{27}$ | $Al_{15}Hf_{25}Nb_{32}Ti_{28}$ | $Al_{10}Hf_{20}Nb_{22}Ti_{33}V_{15}$ |
|-----|-------------------------------|-------------------------------|-------------------------------|-------------------------------------|
| Al  | 24.7 | 20.2 | 14.9 | 9.3  |
| Hf  | 26.0 | 25.2 | 25.4 | 20.8 |
| Nb  | 24.3 | 28.2 | 32.0 | 22.0 |
| Ti  | 25.0 | 26.5 | 27.7 | 33.2 |
| V   | -    | -    | -    | 14.7 |

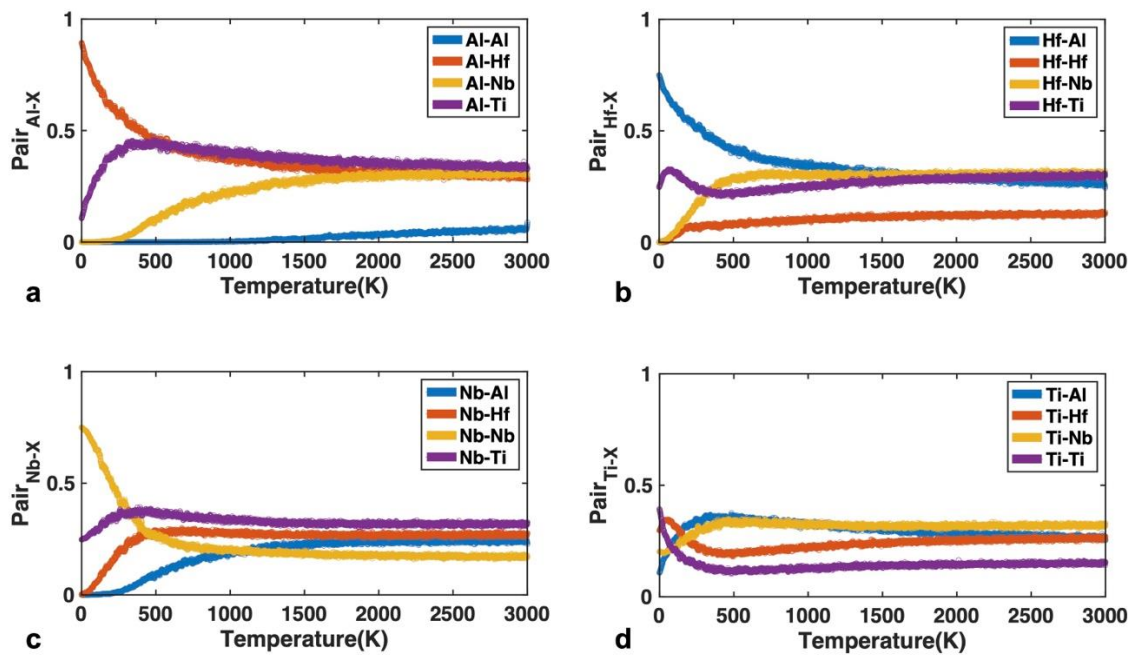

Figure S2. Variation of nearest-neighbor-pair for different elements under various temperatures for $Al_{20}Hf_{24}Nb_{29}Ti_{27}$.

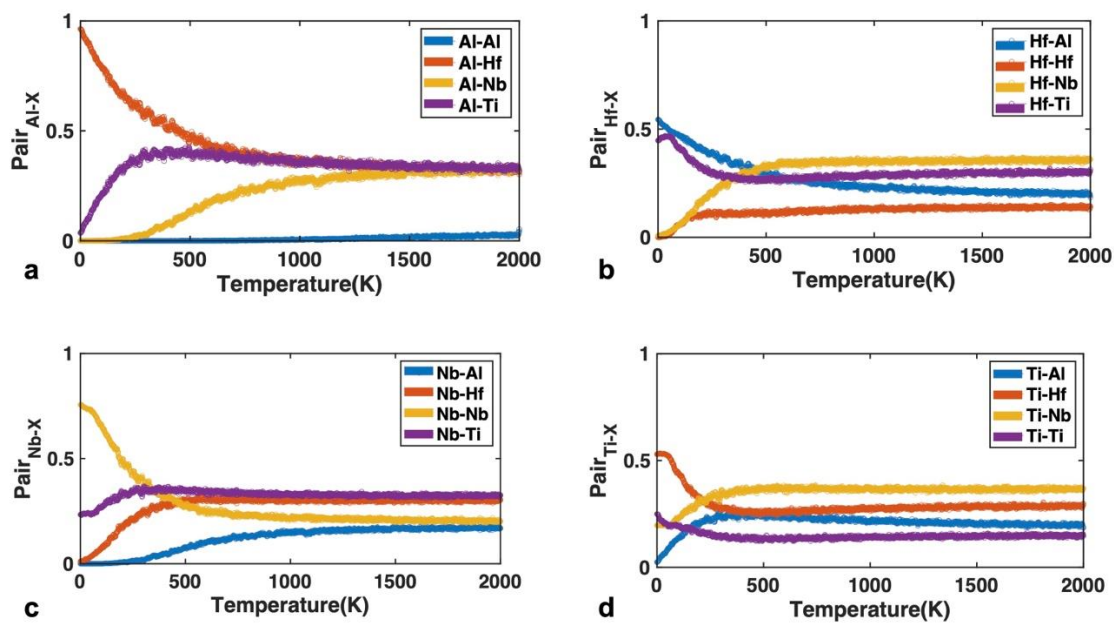

Figure S3. Variation of nearest-neighbor-pair for different elements under various temperatures for $Al_{15}Hf_{25}Nb_{32}Ti_{28}$.

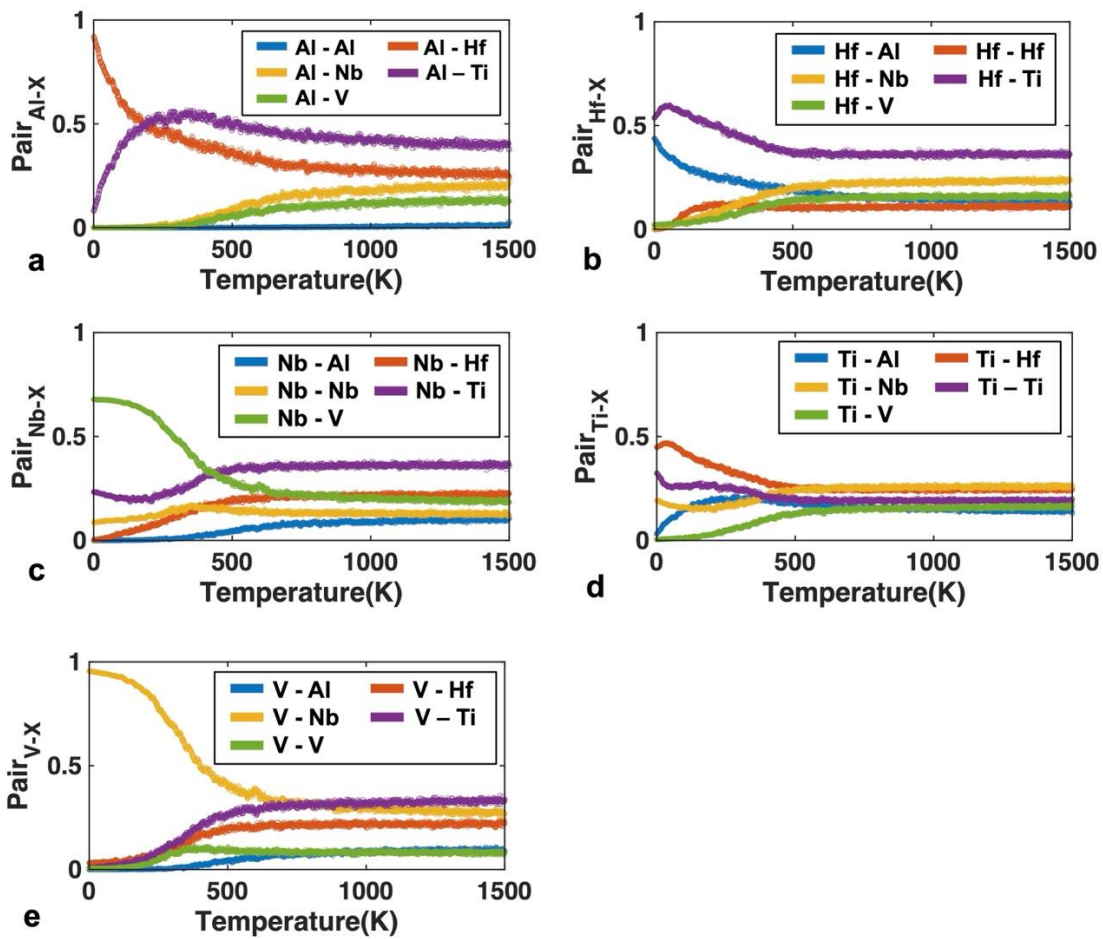

Figure S4. Variation of nearest-neighbor-pair for different elements under various temperatures for $Al_{10}Hf_{20}Nb_{22}Ti_{33}V_{15}$.